\documentclass[10pt,english,draftclsnofoot,english,journal,onecolumn,12pt]{IEEEtran}
\usepackage[T1]{fontenc}
\usepackage[active]{srcltx}
\usepackage{color}
\usepackage{float}
\usepackage{amsthm}
\usepackage{amsmath}
\usepackage{amssymb}
\usepackage{graphicx}

\makeatletter

\floatstyle{ruled}
\newfloat{algorithm}{tbp}{loa}
\providecommand{\algorithmname}{Algorithm}
\floatname{algorithm}{\protect\algorithmname}

  \theoremstyle{definition}
  \newtheorem{defn}{\protect\definitionname}
  \theoremstyle{plain}
  \newtheorem{lem}{\protect\lemmaname}
  \theoremstyle{plain}
  \newtheorem{thm}{\protect\theoremname}

\ifCLASSOPTIONcompsoc
\usepackage[caption=false,font=normalsize,labelfont=sf,textfont=sf]{subfig}
\else
\usepackage[caption=false,font=footnotesize]{subfig}
\fi

\usepackage{cite}
\usepackage{bm}
\usepackage{algorithmic}
\usepackage{algorithm}
\usepackage{graphicx}
\IEEEoverridecommandlockouts

\makeatother

\usepackage{babel}
\providecommand{\definitionname}{Definition}
\providecommand{\lemmaname}{Lemma}
\providecommand{\theoremname}{Theorem}

\begin{document}

\title{\textcolor{black}{Optimal Dynamic Multi-Resource Management in Earth
Observation Oriented Space Information Networks }}

\author{\IEEEauthorblockN{Yu Wang\textsuperscript{}, Min~Sheng\textsuperscript{},~\IEEEmembership{Senior Member,~IEEE},
Qiang Ye,~\IEEEmembership{Member,~IEEE}, \\ Shan Zhang, Weihua
Zhuang\textsuperscript{},~\IEEEmembership{Fellow,~IEEE}, Jiandong
Li\textsuperscript{}, \IEEEmembership{Senior Member,~IEEE}\textsuperscript{}\\}}
\maketitle
\begin{abstract}
Space information network (SIN) is an innovative networking architecture
to achieve near-real-time mass data observation, processing and transmission
over the globe. In the SIN environment, it is essential to coordinate
multi-dimensional heterogeneous resources (i.e., observation resource,
computation resource and transmission resource) to improve network
performance. However, the time varying property of both the observation
resource and transmission resource is not fully exploited in existing
studies. Dynamic resource management according to instantaneous channel
conditions has a potential to enhance network performance. To this
end, in this paper, we study the multi-resource dynamic management
problem, considering stochastic observation and transmission channel
conditions in SINs. Specifically, we develop an aggregate optimization
framework for observation scheduling, compression ratio selection
and transmission scheduling, and formulate a flow optimization problem
based on extended time expanded graph (ETEG) to maximize the sum network
utility. Then, we equivalently transform the flow optimization problem
on ETEG as a queue stability-related stochastic optimization problem.
An online algorithm is proposed to solve the problem in a slot-by-slot
manner by exploiting the Lyapunov optimization technique. Performance
analysis shows that the proposed algorithm achieves close-to-optimal
network utility while guaranteeing bounded queue occupancy. Extensive
simulation results further validate the efficiency of the proposed
algorithm and evaluate the impacts of various network parameters on
the algorithm performance. \end{abstract}

\section{Introduction}

\textcolor{black}{It is well recognized that Earth observation plays
an indispensable role }in realizing a plethora of applications, e.g.,
environment monitoring, weather forecast, target surveillance, and
disaster relief \cite{TSMC16_CoordinatedPlanning,satellite_IoT}.\textcolor{black}{{}
Everyday, huge volumes of data (e.g., some 700 Gbytes of data daily
for ALOS \cite{ALOS}) are generated by versatile applications, which
leads us to the era of big Earth observation data \cite{BigDataforRemoteSensing}.
To accommodate the ever-increasing Earth observation demands, the
space} information network (SIN) \cite{WCM17_ResourceManagement,ComMag_SecurityinSIN,WCM16_CooperativeEarthObservation,WCM_DSCinSpaceInformationNetworks,Mag17_SDSAG}
stands out as a promising solution by integrating multi-layered heterogeneous
space platforms such as geostationary Earth orbit (GEO) satellites,
low Earth orbit (LEO) satellites, high-altitude platforms, and so
on. Through SIN, it is feasible to achieve near-real-time mass data
acquisition, processing and transmission \cite{TM16_ResourceAllocationinSIN}.

In the dynamic and complex SIN environment, multi-resource needs to
be well coordinated for cooperative Earth observation \cite{JSAC18_Multi-ResourceCoordinateScheuling}.
Specifically, to fulfill an Earth observation task, observation, computation
and transmission resources are required. However, the multi-dimensional
heterogeneous resources are normally unbalanced and constrained in
SINs. This constraint can lead to a dilemma that the captured large-volume
observation data cannot be transmitted to specified destinations in
due time. To address the issue, several offline multi-resource coordinate
scheduling algorithms have been proposed in the literature \cite{Ref08_PlanningandSchedulingCOSMO,JSEE16_CoordinateScheduling,CL16_AnalyticalFramework,TCOM17_MissionAwareCPD,JSAC18_Multi-ResourceCoordinateScheuling}.
Particularly, the joint observation and transmission scheduling problem
for the COSMO-SkyMed constellation is first introduced in \cite{Ref08_PlanningandSchedulingCOSMO}.
In \cite{JSEE16_CoordinateScheduling}, a constraint satisfaction
optimization model is used to describe Earth observation satellite
(EOS) observation tasks and data transmission jobs in an integrated
way, and a genetic algorithm based meta-heuristic is proposed. In
our earlier studies, we exploit an extended time expanded graph (ETEG)
method and propose an analytical framework to characterize multi-resource
evolution in the complex and dynamic SIN environment \cite{CL16_AnalyticalFramework,TCOM17_MissionAwareCPD}.
As an extension, multi-resource coordinate scheduling problem is studied
and an iterative optimization technique with low complexity is applied
to efficiently solve the problem in \cite{JSAC18_Multi-ResourceCoordinateScheuling}.

The multi-dimensional resources normally exhibit time varying characteristics
in the SIN context. To be specific, on one hand, for the observation
resource, the resultant image quality depends heavily on the weather
condition of the target area at the time of taking images. An imaging
request for a certain observation area may fail due to bad weather
conditions (raining or with cloud coverage) over that area \cite{TSMC07_ImagingOrderScheduling}.
On the other hand, for the transmission resource, the satellite downlink
contact capacity changes with time due to physical phenomena related
to the propagation of radiowaves through the atmosphere, especially
for satellite communication systems operating at Ku, Ka and V frequency
bands. Further, the capacity of intersatellite contacts also changes
due to varying distance, noise and interference \cite{A.D.Panagopoulos2004,CL16_DynamicContactPlanDesign,PerformanceEvaluation15_AnEfficientUtilization}.
The above observations have given rise to the need for efficient multi-resource
management while considering the time varying property of different
resources.

It is technically challenging to develop dynamic multi-resource coordination
strategies for SINs, due to several reasons: 1) Highly dynamic network
topology \cite{MEO/LEO_TrafficDistribution,predictable_routing2}.
The continuously changing network topology brings a resource availability
issue. A certain type of resource (e.g., observation resource) is
usable only when two related nodes are within the line-of-sight range.
Besides, the resource combinations to accomplish a task are highly
complex. As a consequence, it is difficult to model the multi-resource
correlation relationship in the presence of topology dynamics; 2)
Multi-resource conflicts \cite{TwoPhaseScheduling,TWC16_CollaborativeDataDownloading,Mag_ContactDesignChallenge}.
Both the observation and transmission resources are limited in SINs.
These multi-resource limitations can lead to infeasibility of all
potential observation and transmission opportunities, and thus induce
observation and transmission conflicts. \textcolor{black}{To make
matter worse}, the above conflicts can change in a small time period
due to the dynamic network topology evolution; and 3) Balancing the
stochastic observation and transmission processes. Since both the
observation process and transmission process exhibit time varying
characteristics, it is difficult to match the two stochastic processes
without prior knowledge of channel state distribution.

In this paper, to address the technical challenges, we study the multi-dimensional
resource dynamic management problem for SINs, while taking into account
both the observation and transmission channel variations. Specifically,
we first employ an ETEG to characterize multi-resource correlation
relationships in the dynamic SIN context. On the basis of ETEG, an
aggregate optimization framework is developed for observation scheduling,
compression ratio selection and transmission scheduling, and a constrained
flow optimization problem is formulated to maximize the sum network
utility. We then transform the optimization problem on ETEG as a queue
stability-related stochastic optimization problem on a multi-queue
multi-server queueing model. By exploiting the Lyapunov optimization
technique, an online algorithm, i.e., Dynamic Multi-Resource Cooperation
(DMRC) algorithm, is proposed to decompose the optimization problem
into separate \textcolor{black}{joint observation scheduling and adaptive
processing subproblem} and transmission scheduling sub-problem. For
the  first sub-problem, we utilize the Lagrangian dual theory to optimally
solve the transformed convex optimization problem. For the second
sub-problem, we transform it into the maximum weighted matching problem
in a constructed bipartite graph. The computational complexity and
theoretical performance bound of the proposed DMRC algorithm are analyzed
in detail. We show that the DMRC algorithm can achieve network utility
arbitrarily close to the optimal value without requiring the prior
knowledge of channel conditions. Extensive simulation results are
provided to demonstrate the efficiency of the proposed algorithm and
evaluate the impacts of various network parameters on the algorithm
performance.

In a nutshell, the main contributions of this paper are summarized
as follows:
\begin{enumerate}
\item An optimization framework of observation scheduling, adaptive compression
and transmission scheduling based on ETEG is formulated to maximize
the sum network utility;
\item We equivalently transform the constrained flow optimization problem
on ETEG as a queue stability-related stochastic optimization problem
by exploiting a multi-queue multi-server queueing model;
\item We use the Lyapunov optimization method to solve the optimization
problem. The performance bound of the proposed DMRC algorithm is theoretically
analyzed and proved to be arbitrarily close to the optimality.
\end{enumerate}

The remainder of this paper is organized as follows. Section \ref{sec:System-Model}
introduces the SIN system model under consideration and Section \ref{sec:Problem Formulation}
elaborates on the detailed problem formulation and transformation.
We propose an approximate online multi-resource dynamic scheduling
algorithm and analyze its complexity as well as performance bound
in Section \ref{sec:Proposed-Solution-and-Performance Analysis}.
The performance evaluation by simulations is presented in Section
\ref{sec:Performance-Evaluation}, followed by concluding remarks
and future research in Section \ref{sec:Conclusion}.

\section{System Model\label{sec:System-Model} }

In this section, we first introduce the SIN network model under consideration.
Then, an ETEG model is employed to describe the multi-resource correlation
relationship over the network.

\subsection{Network Model}

\begin{figure*}
\centering\includegraphics{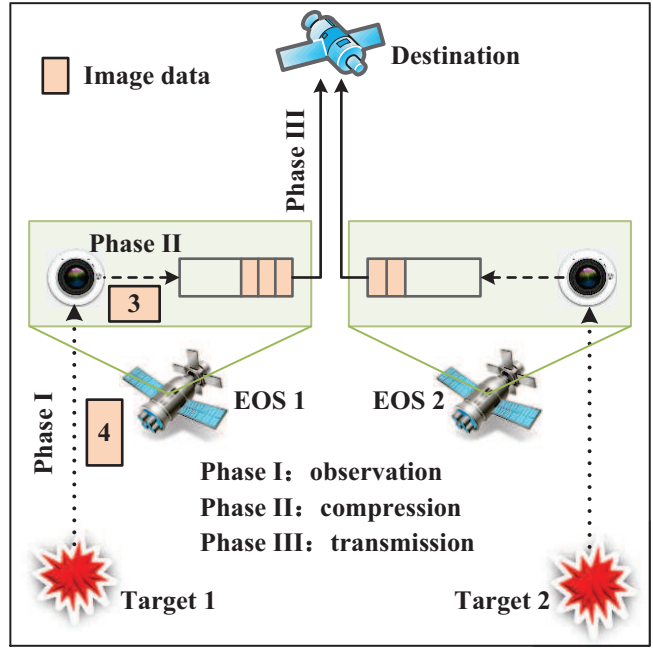}\vspace{-1.0em}

\caption{An example SIN system with 2 targets, 2 EOSs and a destination.\label{fig:An-example-SIN}}
\end{figure*}

Consider a satellite-based SIN system operating in slotted time $t\in\mathcal{T}=\{1,2,...\}$
\cite{ResourceAllocationSpaceSystem}. The length of a time slot is
denoted as $\tau$. There are three different types of components
in the SIN: 1) a set, $\mathcal{I}=\{1,2,\ldots,I\}$, of $I$ ground
targets that need to be continuously monitored; 2) a set, $\mathcal{K}=\{1,\ldots,K\}$,
of $K$ EOSs moving in the LEOs to acquire observation images from
the targets of interest, perform adaptive compression (i.e., select
an appropriate compression ratio for the raw image data), and then
transmit those compressed data to specified destinations; and 3) a
set, $\mathcal{N}=\{1,\ldots,N\}$, of $N$ destinations (i.e., relay
satellites or ground stations) which serve as the sinks for all the
observation data. There are $I$ flows in the network, with each flow
corresponding to end-to-end service for a target. An example SIN system
with 2 targets, 2 EOSs and a destination is shown in Fig. \ref{fig:An-example-SIN}.

As can be seen in Fig. \ref{fig:An-example-SIN}, it encompasses three
phases, namely observation phase, compression phase and transmission
phase, to serve a flow. Firstly, in the observation phase, an EOS
is scheduled to collect observation data (i.e., images) using its
onboard imaging camera when it is in the line-of-sight of the associated
target. Secondly, in the compression phase, the EOS performs necessary
processing and selects an appropriate compression ratio for the collected
raw image data. For example, EOS 1 adopts $\frac{3}{4}$ compression
ratio%
\footnote{In this paper, the compression ratio is defined as data volume after
image compression to raw image data volume.%
} to process the image data. The compressed observation data are then
stored in an onboard buffer. Thirdly, in the transmission phase, the
EOS uses the store-and-forward method \cite{topology_control2} to
download the data to a destination after it enters the coverage of
the destination.

\subsection{Graph Model}

\subsubsection{The Original ETEG Model}

We employ an ETEG \cite{Report_RoutinginSpaceandTime,TVT16_V2VForwardingforGreen}
$\mathcal{G}=(\mathcal{V}_{t},\mathcal{E}_{t},\mathcal{T})$ to reveal
the multi-resource correlation relationship over the dynamic SIN context
during the whole time horizon $\mathcal{T}$, where $\mathcal{V}_{t}$
and $\mathcal{E}_{t}$ denote the set of the vertices and edges at
time slot $t$, respectively. The detailed construction procedure
of ETEG is given as follows.
\begin{figure*}
\centering\includegraphics[scale=0.5]{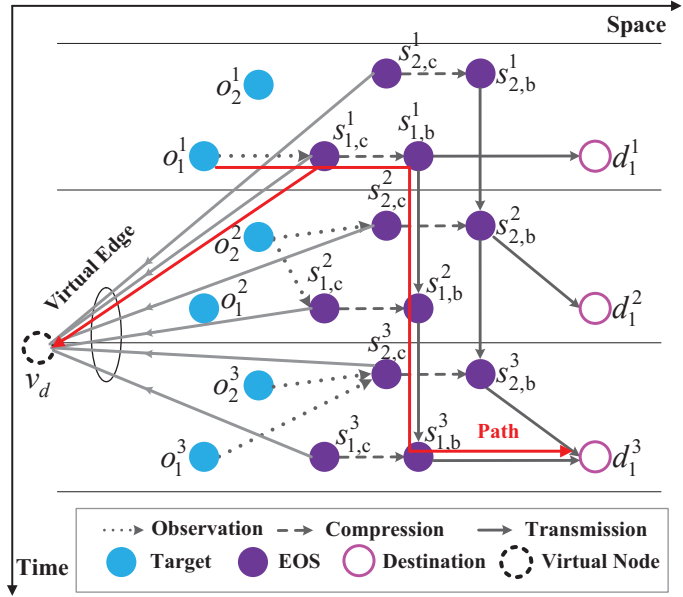}\vspace{-1.0em}

\caption{Corresponding ETEG for the example SIN. (b) ETEG transformation.\label{fig:ETEG}}
\end{figure*}

\textit{i) Vertices:}\textsl{ }The ETEG consists of $T$ layers. Each
layer captures the network status (i.e., network topology and channel
condition) at a single time slot. The network status remains constant
at a time slot, and can instantaneously change during slot transitions.
Specifically, at time slot $t$ $(1\leq t\leq T)$, target $i\in\mathcal{I}$,
EOS $k\in\mathcal{K}$, and destination $n\in\mathcal{N}$ are represented
by vertices $o_{i}^{t}$, $s_{k}^{t}$, and $d_{n}^{t}$ in the ETEG,
respectively. Besides, we divide an EOS-related vertex, $s_{k}^{t}$,
into two components, namely imaging camera, $s_{k,\textrm{c}}^{t}$,
and buffer, $s_{k,\textrm{b}}^{t}$. This division aims to clearly
reflect the three phases to accomplish a flow as to be described.

\textit{ii) Edges:}\textsl{ }There are three different types of edges
in $\mathcal{E}_{t}$, i.e., observation edge $\mathcal{E}_{t}^{\textrm{ob}}$,
compression edge $\mathcal{E}_{t}^{\textrm{co}}$, and transmission
edge $\mathcal{E}_{t}^{\textrm{tr}}$. Each type of edge describes
a related phase, e.g., observation edge corresponds to the observation
phase. In particular, an observation edge $(o_{i}^{t},s_{k,\textrm{c}}^{t})\in\mathcal{E}_{t}^{\textrm{ob}}$
exists, if an observation opportunity is present between target $i$
and EOS $k$ at time slot $t$. The capacity of observation edge $(o_{i}^{t},s_{k,\textrm{c}}^{t})$
is denoted by $B_{i,k}(t)$. Therefore, at time slot $t$, the stochastic
observation channel state is expressed by an $I\times K$ matrix $\boldsymbol{B}(t)=\left\{ B_{i,k}(t)|(o_{i}^{t},s_{k,\textrm{c}}^{t})\in\mathcal{E}_{t}^{\textrm{ob}}\right\} .$
Denote $x_{i,k}(t)$ as the observation scheduling function, where
$x_{i,k}(t)=1$ indicates that observation edge $(o_{i}^{t},s_{k,\textrm{c}}^{t})$
is selected for observation at time slot $t$, and $x_{i,k}(t)=0$
otherwise. Subsequently, we can compute the amount of observation
data volume $w(o_{i}^{t},s_{k,\textrm{c}}^{t})$ acquired by edge
$(o_{i}^{t},s_{k,\textrm{c}}^{t})$ as $w(o_{i}^{t},s_{k,\textrm{c}}^{t})=B_{i,k}(t)\cdot x_{i,k}(t)\cdot\tau$.
On the other hand, regarding EOS $k$, a compression edge $(s_{k,\textrm{c}}^{t},s_{k,\textrm{b}}^{t})\in\mathcal{E}_{t}^{\textrm{co}}$
exists at time slot $t$. Denote $w(s_{k,\textrm{c}}^{t},s_{k,\textrm{b}}^{t})$
as the data volume of edge $(s_{k,\textrm{c}}^{t},s_{k,\textrm{b}}^{t})$
after image compression. To distinguish the served flow on edge $(s_{k,\textrm{c}}^{t},s_{k,\textrm{b}}^{t})$,
we further denote $w(s_{k,\textrm{c}}^{t},s_{k,\textrm{b}}^{t},i)$
as the data volume for flow $i$, i.e., target $i$.

In parallel, the transmission edge can be further separated into a
forward-transmission edge and a store-transmission edge, taking into
account the store-and-forward transmission pattern in SIN. Specifically,
forward-transmission edge $(s_{k,\textrm{b}}^{t},d_{n}^{t})\in\mathcal{E}_{t}^{\textrm{tr}}$
exists, if a transmission opportunity between EOS $k$ and destination
$n$ is available during the time slot. The capacity of forward-transmission
edge $(s_{k,\textrm{b}}^{t},d_{n}^{t})$ is denoted by $C_{k,n}(t)$.
Therefore, at time slot $t$, the stochastic transmission channel
state is expressed by a $K\times N$ matrix $\boldsymbol{C}(t)=\left\{ C_{k,n}(t)|(s_{k,\textrm{b}}^{t},d_{n}^{t})\in\mathcal{E}_{t}^{\textrm{tr}}\right\} .$
Denote $y_{k,n}(t)$ as the transmission scheduling function, where
$y_{k,n}(t)=1$ indicates that forward-transmission edge $(s_{k,\textrm{b}}^{t},d_{n}^{t})$
is selected for data transmission at time slot $t$, and $y_{k,n}(t)=0$
otherwise. Let $w(s_{k,\textrm{b}}^{t},d_{n}^{t},i)$ be the amount
of data volume delivered by forward-transmission edge $(s_{k,\textrm{b}}^{t},d_{n}^{t})$
for flow $i$. Meanwhile, store-transmission edge $(s_{k,\textrm{b}}^{t},s_{k,\textrm{b}}^{t+1})\in\mathcal{E}_{t}^{\textrm{tr}}$
is used to model that EOS $k$ can physically carry its data forward
from time slot $t$ to time slot $t+1$. Denote $w(s_{k,\textrm{b}}^{t},s_{k,\textrm{b}}^{t+1})$
as the stored data volume of edge $(s_{k,\textrm{b}}^{t},s_{k,\textrm{b}}^{t+1})$.
Likewise, we further denote $w(s_{k,\textrm{b}}^{t},s_{k,\textrm{b}}^{t+1},i)$
as the amount of stored data volume for flow $i$. Since an EOS typically
can have mass data storage capacity, we assume that each EOS has infinite
buffer size.

\textit{iii) Paths: }A path is an ordered vertex sequence in the ETEG.
Note that a path originates from a vertex representing the target
and terminates at a vertex representing the destination. The set of
edges that it traverses can thus capture the sequentially chained
multi-dimensional resources. For example, path $\{o_{1}^{1},s_{1,\textrm{c}}^{1},s_{1,\textrm{b}}^{1},s_{1,\textrm{b}}^{2},s_{1,\textrm{b}}^{3},d_{1}^{3}\}$,
represented by the red line in Fig. \ref{fig:ETEG}, shows that at
the first time slot, target $o_{1}^{1}$ is observed by EOS $s_{1,\textrm{c}}^{1}$,
the acquired image data are compressed and then stored on $s_{1,\textrm{b}}^{1}$
for 2 time slots, and finally at the third time slot, the data are
transmitted to destination $d_{1}^{3}$. In order to satisfy flow
conservation, we add a virtual node $v_{d}$ and the corresponding
virtual edges $\{(s_{k,\textrm{c}}^{t},v_{d})\}$ in the original
ETEG. During the compression phase, a portion of raw image data flow
away from $s_{k,\textrm{c}}^{t}$ owing to data compression. The virtual
edge, $(s_{k,\textrm{c}}^{t},v_{d})$, is to make up for the compressed
data by compression edge $(s_{k,\textrm{c}}^{t},s_{k,\textrm{b}}^{t})$.

\subsubsection{ETEG Transformation}

\begin{figure*}
\centering\includegraphics[scale=0.5]{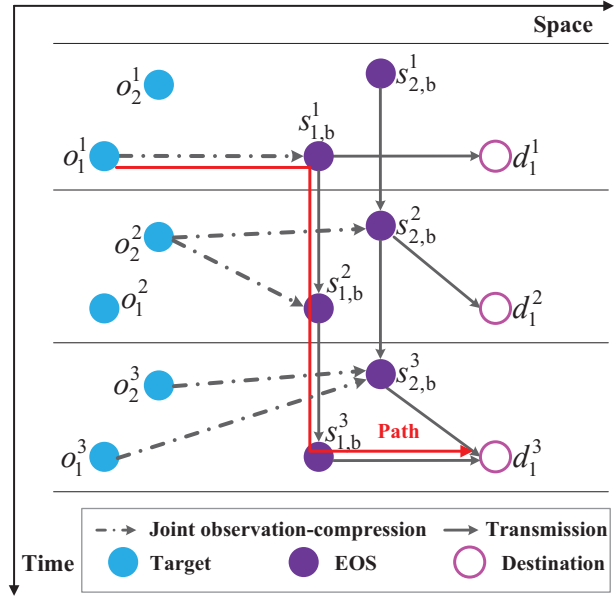}\vspace{-1.0em}

\caption{ETEG transformation.\label{fig:ETEG-trans}}
\end{figure*}

To reduce the complexity of constructing ETEG, we transform the initial
graph in two steps: 1) to combine observation edge and compression
edge as an integrated edge, namely joint observation-compression (JOC)
edge. We use a JOC edge $(o_{i}^{t},s_{k,\textrm{b}}^{t})$ to replace
the observation edge $(o_{i}^{t},s_{k,\textrm{c}}^{t})$ and compression
edge $(s_{k,\textrm{c}}^{t},s_{k,\textrm{b}}^{t})$. Denote the set
of JOC edges at time slot $t$ as $\mathcal{E}_{t}^{\textrm{joc}}$.
Naturally, the data volume $w(o_{i}^{t},s_{k,\textrm{b}}^{t})$ of
JOC edge $(o_{i}^{t},s_{k,\textrm{b}}^{t})$ equals to $x_{i,k}(t)B_{i,k}(t)\varrho_{k}^{i}(t)$,
where $\varrho_{k}^{i}(t)$ represents the image compressing ratio
for flow $i$ at time slot $t$; and 2) to remove the virtual node
$v_{d}$ and all associated virtual edges $\{(s_{k,\textrm{c}}^{t},v_{d})\}$.
In this case, the flow conservation constraint can be guaranteed without
introducing additional virtual node and edges, since \textcolor{black}{the
compressed data are already excluded} by the JOC edges. The corresponding
graph after transformation is given in Fig. \ref{fig:ETEG-trans}.
Clearly, the modified graph is simpler.

On the basis of the transformed ETEG, the flow conservation constraint
at an EOS can be expressed as: \vspace{-1.0em}

\begin{equation}
x_{i,k}(t)w(s_{k,\textrm{c}}^{t},s_{k,\textrm{b}}^{t},i)+w(s_{k,\textrm{b}}^{t-1},s_{k,\textrm{b}}^{t},i)=w(s_{k,\textrm{b}}^{t},s_{k,\textrm{b}}^{t+1},i)+\sum_{d_{n}^{t}:(s_{k,\textrm{b}}^{t},d_{n}^{t})\in\mathcal{E}_{t}^{\textrm{tr}}}y_{k,n}(t)w(s_{k,\textrm{b}}^{t},d_{n}^{t},i)\,\,\,\forall i,k,t.\label{eq:flow_conservation}
\end{equation}
To keep notation succinct, we set $w(s_{k,\textrm{b}}^{0},s_{k,\textrm{b}}^{1},i)=0$
and $w(s_{k,\textrm{b}}^{T},s_{k,\textrm{b}}^{T+1},i)=0$. It can
be seen that, for each flow $i$ on an EOS $k$, the total observation
data volume after compression plus the stored data from time slot
$t-1$ should equal to that delivered to destinations plus the stored
data to time slot $t+1$.

\section{Problem Formulation\label{sec:Problem Formulation}}

In this section, based on ETEG, the network utility maximization problem
is formulated. After that, we reformulate the problem by taking advantage
of a queueing model.

\subsection{Optimization Problem}

\textcolor{black}{Given a constructed ETEG, the problem under consideration
is to jointly decide observation scheduling, transmission scheduling
and compression ratio selection. Our objective is to maximize the
}sum long-term average network utility\textcolor{black}{. }Note that
utility function $U_{i}(A_{i}(t))$ should be increasing, continuously
differentiable and strictly concave in $A_{i}(t)$. The concavity
of the utility function is based on that the marginal utility decreases
as the amount of collected observation data increases in SIN\textcolor{black}{.
Without loss of generality, we define the network utility function}%
\footnote{\textcolor{black}{Other network utility functions can be applied as
well, and the following analysis can be extended in a similar way. }%
}\textcolor{black}{{} as $U_{i}(A_{k}^{i}(t))=\log(1+A_{k}^{i}(t)),\:\forall k,i$
\cite{JSAC16_UtilityOptimalResourceManagement,[JSAC11]_DelayAwareCrossLayerDesignforNUM}.
We formulate it as optimization problem (P1):}\vspace{-1.0em}

\begin{eqnarray}
(\textrm{P1}) & \max_{\boldsymbol{X},\boldsymbol{Y},\boldsymbol{w},\boldsymbol{\varrho}} & \lim_{t\rightarrow\infty}\frac{1}{T}\sum_{t=1}^{T}\sum_{i=1}^{I}U_{i}(A_{i}(t))\nonumber \\
 & s.t. & \textrm{C1}:\:\textrm{Flow conservation constraint (\ref{eq:flow_conservation})}\nonumber \\
 &  & \textrm{C2}:\:\sum_{i}x_{i,k}(t)\leq1\,\,\,\forall k,t\nonumber \\
 &  & \textrm{C3}:\:\sum_{k}x_{i,k}(t)\leq1\,\,\,\forall i,t\nonumber \\
 &  & \textrm{C4}:\:\sum_{n}y_{k,n}(t)\leq1\,\,\,\forall k,t\nonumber \\
 &  & \textrm{C5}:\:\sum_{k}y_{k,n}(t)\leq M_{n}\,\,\,\forall n,t\nonumber \\
 &  & \textrm{C6}:\:\sum_{i}w(s_{k,\textrm{b}}^{t},d_{n}^{t},i)\leq y_{k,n}(t)C_{k,n}(t)\tau\,\,\,\forall(s_{k,\textrm{b}}^{t},d_{n}^{t})\in\mathcal{E}_{t}^{\textrm{tr}},t\nonumber \\
 &  & \textrm{C7}:\:\overline{A}_{i}\geq a_{i}\,\,\,\forall i\nonumber \\
 &  & \textrm{C8}:\: y_{k,n}(t)\in\{0,1\}\,\,\,\forall(s_{k,\textrm{b}}^{t},d_{n}^{t})\in\mathcal{E}_{t}^{\textrm{tr}},t\nonumber \\
 &  & \textrm{C9}:\: x_{i,k}(t)\in\{0,1\}\,\,\,\forall(o_{i}^{t},s_{k,\textrm{b}}^{t})\in\mathcal{E}_{t}^{\textrm{joc}},t\nonumber \\
 &  & \textrm{C10}:\:\varrho_{k}^{i}(t)\in\Lambda=\{\varrho_{1},...,\varrho_{J}\}\,\,\,\forall(o_{i}^{t},s_{k,\textrm{b}}^{t})\in\mathcal{E}_{t}^{\textrm{joc}},t\label{eq:optimization problem}
\end{eqnarray}
where $X=\{x_{i,k}(t)\}$ and $Y=\{y_{k,n}(t)\}$ are the observation
scheduling matrix and transmission scheduling matrix, respectively,
$\boldsymbol{w}=\{w(s_{k,\textrm{c}}^{t},s_{k,\textrm{b}}^{t},i)\}$
is the link capacity allocation matrix, and $\boldsymbol{\varrho}=\{\varrho_{k}^{i}(t)\}$
is the compression ratio selection matrix. In problem (P1), C1 ensures
flow conservation for all the EOSs. C2 states that each EOS can observe
at most one target at a time slot, while C3 states that each target
can be observed at most once at a time slot to reduce data redundancy.
C4 specifies that each EOS can transmit to at most one destination
at a time slot. Similarly, C5 implies that destination $n$ can support
at most $M_{n}$ concurrent transmissions at a time slot. C6 is the
transmission link capacity constraint. C7 indicates that the long-term
average rate $\overline{A}_{i}$ for flow $i$ is not smaller than
a pre-defined threshold $a_{i}$. C8 and C9 correspond to the binary
scheduling variables $x_{i,k}(t)$ and $y_{k,n}(t)$, respectively.
C10 restricts the compression ratio to a predefined set $\Lambda=\{\varrho_{1},...,\varrho_{J}\}$,
where $\varrho_{j}$ is the $j$-th compression ratio and $\varrho_{J}\leq\cdots\leq\varrho_{j}\leq\varrho_{1}\leq1$.

It can be seen that problem (P1) is a mixed integer linear programming
(MILP) problem, which is generally NP-hard in nature \cite{book_ApproximationAlgorithm}.
Moreover, constraints C1 and C7 introduce complex coupling relationships
among multiple time slots \cite{dtn_graph}. This exacerbates the
computational complexity to solve the problem. Last but not the least,
both the observation and transmission channel conditions are stochastic
and cannot be predicted in advance. In this regard, it is challenging
to maximize network performance without prior knowledge of channel
quality.

\subsection{\textcolor{black}{Problem Transformation}}

\begin{figure}
\centering\includegraphics[scale=1.2]{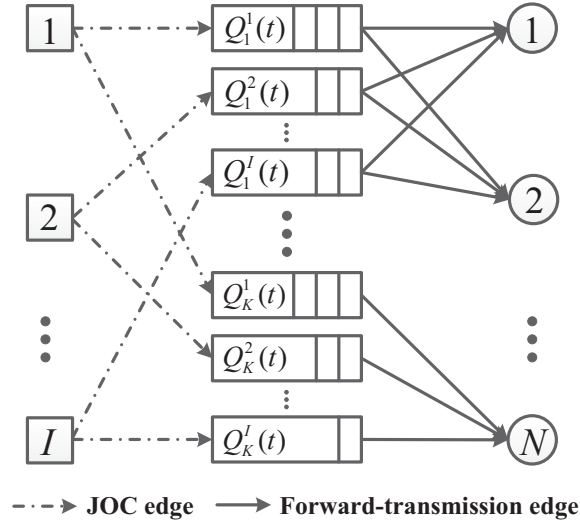}\vspace{-1.0em}\caption{Multi-queue multi-server queueing model for the concerned SIN system
model.\label{fig:Multi-queue-model}}
\end{figure}

Herein, as shown in Fig. \ref{fig:Multi-queue-model}, we present
a multi-queue multi-server queueing model to reformulate problem (P1).
Let $Q_{k}^{i}(t)$ denote the data queue occupancy of EOS $k$ for
flow $i$ at time slot $t$ and $\boldsymbol{Q}(t)=(Q_{1}^{1}(t),Q_{1}^{2}(t),\ldots,Q_{K}^{I}(t))$
represent the vector of data queues for all EOSs. Note that $Q_{k}^{i}(t)$
equals to the data volume $w(s_{k,\textrm{b}}^{t-1},s_{k,\textrm{b}}^{t},i)$
of store-transmission edge $(s_{k,\textrm{b}}^{t-1},s_{k,\textrm{b}}^{t})$.
In addition, the data volume carried by a JOC edge (or forward-transmission
edge) reflects the arrival process (or service process) to a data
queue. Thus, the dynamics of a data queue $Q_{k}^{i}(t+1)$ can be
expressed as
 \vspace{-1.5em}

\begin{equation}
Q_{k}^{i}(t+1)=[Q_{k}^{i}(t)-\mu_{k}^{i}(t)]^{+}+A_{k}^{i}(t),\:\forall k,i\label{eq:queue_dynamic}
\end{equation}
where $\mu_{k}^{i}(t)=\sum_{d_{n}^{t}:(s_{k,\textrm{b}}^{t},d_{n}^{t})\in\mathcal{E}_{t}^{\textrm{tr}}}y_{k,n}(t)w(s_{k,\textrm{b}}^{t},d_{n}^{t},i)$
captures the service process. Meanwhile, $A_{k}^{i}(t)=\varrho_{k}^{i}(t)x_{i,k}(t)B_{i,k}(t)$
models the equivalent\textcolor{black}{{} traffic} arrival process.
We further denote $A_{i}(t)$ as the sum rate for flow $i$ at time
slot $t$, i.e., $A_{i}(t)=\sum_{k}A_{k}^{i}(t)$. Besides, $Z^{+}=\max(0,Z)$
holds. In the following, we give a useful definition of queue stability.
\begin{defn}
\textcolor{black}{Queue $Q(t)$ is mean rate stable if $\underset{t\rightarrow\infty}{\lim}\frac{1}{t}\mathbb{E}[Q(t)]=0$
holds. A network of queues is stable if all the individual queues
are mean rate stable.}
\end{defn}
According to the definition, we can gain the following two insights:
1) network stability is guaranteed if the length of all queues is
finite; and 2) applying Little\textquoteright{}s Theorem, we can depict
average delay by queue length and further by queue stability for a
stable queueing system.
\begin{lem}
\textup{\textcolor{black}{Let queue $Q(t)$ be associated with stochastic
arrival and service processes $A(t)$ and $u(t)$. The time averages
of both processes converge to $\overline{A}$ and $\overline{\mu}$,
respectively. The queue, $Q(t)$, is mean rate stable if and only
if $\overline{\mu}\geq\overline{A}$ holds. }}
\end{lem}
The proof of Lemma 1 can be found in \cite{book_StochasticNetworkOptimization}.
The intuition behind the lemma is that the network is stable if the
traffic arrival can be served by the network in the long-term. As
a result, the backlog of all queues is finite.
\begin{thm}
\label{thm:queue_stability}C1 can be approximately transformed into
the mean rate queue stability constraint as \textup{$\textrm{C0}:\:\underset{t\rightarrow\infty}{\lim}\frac{1}{t}\mathbb{E}\left(|Q_{k}^{i}(t)|\right)=0\,\,\,\forall k,i.$
\vspace{-1.0em}}\end{thm}
\begin{IEEEproof}
Define $\overline{A_{k}^{i}}$ and $\overline{\mu_{k}^{i}}$ as
\begin{eqnarray}
 &  & \overline{A_{k}^{i}}\triangleq\lim_{T\rightarrow\infty}\sum_{t=1}^{T}\frac{1}{T}\left(\varrho_{k}^{i}(t)x_{i,k}(t)B_{i,k}(t)\right)\\
 &  & \overline{\mu_{k}^{i}}\triangleq\lim_{T\rightarrow\infty}\frac{1}{T}\sum_{t=1}^{T}\left(\sum_{d_{n}^{t}:(s_{k,\textrm{b}}^{t},d_{n}^{t})\in\mathcal{E}_{t}^{\textrm{tr}}}y_{k,n}(t)w(s_{k,\textrm{b}}^{t},d_{n}^{t},i)\right).
\end{eqnarray}
For the flow conservation constraint in (\ref{eq:flow_conservation}),
summing over all time slots, dividing it by $T$, and taking the limit,
we obtain \vspace{-1.5em}

\begin{eqnarray}
 & \lim_{T\rightarrow\infty}\sum_{t=1}^{T}\frac{1}{T}\left(\varrho_{k}^{i}(t)x_{i,k}(t)B_{i,k}(t)+w(s_{k,\textrm{b}}^{t-1},s_{k,\textrm{b}}^{t},i)\right)\nonumber \\
= & \lim_{T\rightarrow\infty}\sum_{t=1}^{T}\frac{1}{T}\left(\sum_{d_{n}^{t}:(s_{k,\textrm{b}}^{t},d_{n}^{t})\in\mathcal{E}_{t}^{\textrm{tr}}}y_{k,n}(t)w(s_{k,\textrm{b}}^{t},d_{n}^{t},i)+w(s_{k,\textrm{b}}^{t},s_{k,\textrm{b}}^{t+1},i)\right).
\end{eqnarray}
Recall that $w(s_{k,\textrm{b}}^{0},s_{k,\textrm{b}}^{1},i)=w(s_{k,\textrm{b}}^{T},s_{k,\textrm{b}}^{T+1},i)=0$
holds, it can thus be seen $\sum_{t=1}^{T}w(s_{k,\textrm{b}}^{t-1},s_{k,\textrm{b}}^{t},i)=\sum_{t=1}^{T}w(s_{k,\textrm{b}}^{t},s_{k,\textrm{b}}^{t+1},i)$.
Combining (4), (5) and (6), we obtain $\overline{A_{k}^{i}}=\overline{\mu_{k}^{i}}$.
Since our objective is to maximize a concave function of $A_{i}(t)$,
we can relax the above equation as $\overline{A_{k}^{i}}\leq\overline{\mu_{k}^{i}}$
without trading optimality. As a result, all queues in the network
are mean rate stable, and C0 can be accordingly derived following
Definition 1.
\end{IEEEproof}
According to Theorem \ref{thm:queue_stability}, problem (P1) can
be reformulated as \vspace{-1.5em}

\begin{eqnarray*}
(\textrm{P2}) & \max_{\boldsymbol{X},\boldsymbol{Y},\boldsymbol{w},\boldsymbol{\varrho}} & \lim_{t\rightarrow\infty}\frac{1}{T}\sum_{t=1}^{T}\sum_{i=1}^{I}U_{i}(A_{i}(t))\\
 & s.t. & \textrm{C0, C2-C10}.
\end{eqnarray*}

\section{Proposed Solution and Performance Analysis\label{sec:Proposed-Solution-and-Performance Analysis}}

In this section, we first present an online algorithm, i.e., DMRC
algorithm, to solve problem (P2). After that, we theoretically analyze
its complexity as well as performance.

\subsection{Dynamic Multi-Resource Cooperation Algorithm}

We use the Lyapunov optimization theory \cite{[JSAC11]_DelayAwareCrossLayerDesignforNUM,M.J.Neely2003}
to design an effective online algorithm to solve problem (P2). Firstly,
we define a set of virtual queues $\boldsymbol{P}(t)=(P_{1}(t),P_{2}(t),\ldots,P_{I}(t))$,
with $P_{i}(t)$ being the virtual queue for flow $i$. The virtual
queue $P_{i}(t)$ evolves as \vspace{-1.0em}

\begin{equation}
P_{i}(t+1)=[P_{i}(t)+p_{i}(t)]^{+}\label{eq:virtual queue}
\end{equation}
where $p_{i}(t)=a_{i}-A_{i}(t)$. When all virtual queues, $\boldsymbol{P}(t)$,
are stable, $\overline{p}_{i}\leq0$ holds according to Lemma 1. Consequently,
C7 is automatically satisfied. Let $\boldsymbol{\Theta}(t)=[\boldsymbol{Q}(t),\boldsymbol{P}(t)]$
be a concatenated vector of all actual and virtual queues and define
the quadratic Lyapunov function as\vspace{-1.0em}

\begin{equation}
L(\boldsymbol{\Theta}(t))=\frac{1}{2}\sum_{i=1}^{I}\sum_{k=1}^{K}Q_{k}^{i}(t)^{2}+\frac{1}{2}\sum_{i=1}^{I}P_{i}(t)^{2}.\label{eq:Lyapunov Function}
\end{equation}

Then, the one-slot conditional Lyapunov drift is given by\vspace{-1.0em}

\begin{equation}
\Delta(\boldsymbol{\Theta}(t))=\mathbb{E}\{L(\boldsymbol{\Theta}(t+1))-L(\boldsymbol{\Theta}(t))|\boldsymbol{\Theta}(t)\}.\label{eq:Lypunov Drift}
\end{equation}

\begin{lem}
\textup{Assume} \textup{$\boldsymbol{B}(t)$ and $\boldsymbol{C}(t)$
are independent identically distributed (i.i.d) over time slot $t$.
Under any feasible scheduling policy that satisfies all the constraints
in problem (P1), we have the following inequality:\vspace{-1.0em}}
\end{lem}
\[
\Delta(\boldsymbol{\Theta}(t))-V\sum_{i}\mathbb{E}\left\{ U_{i}(A_{i}(t))|\boldsymbol{\Theta}(t)\right\} \leq\Gamma-V\sum_{i=1}^{I}\mathbb{E}\left\{ U_{i}(A_{i}(t))|\boldsymbol{\Theta}(t)\right\} \vspace{-1.0em}
\]

\begin{equation}
+\sum_{i=1}^{I}\sum_{k=1}^{K}\mathbb{E}\left\{ Q_{k}^{i}(t)(A_{k}^{i}(t)-\mu_{k}^{i}(t))|\boldsymbol{\Theta}(t)\right\} +\sum_{i=1}^{I}\mathbb{E}\left\{ P_{i}(t)(a_{i}-A_{i}(t))|\boldsymbol{\Theta}(t)\right\} \label{eq:drift-plus-penalty}
\end{equation}
where $V$ is a control factor to strike a balance between the queue
backlog and the network utility, and $\Gamma$ is a finite constant
that satisfies\vspace{-1.0em}

\begin{equation}
\Gamma\geq\frac{1}{2}\sum_{i=1}^{I}\sum_{k=1}^{K}\{A_{k}^{i}(t)^{2}+\mu_{k}^{i}(t)^{2}\}+\frac{1}{2}\sum_{i=1}^{I}\{a_{i}-A_{i}(t)\}^{2}.\label{eq:Constant B}
\end{equation}

\begin{IEEEproof}
See Appendix A.
\end{IEEEproof}
The fundamental design philosophy of the proposed DMRC algorithm is
to minimize the right-hand-side of (\ref{eq:drift-plus-penalty})
in each time slot. To be specific, firstly, at every time slot, observe
the data queues $\boldsymbol{Q}(t)$, virtual queues $\boldsymbol{P}(t)$,
and channel conditions $\boldsymbol{B}(t)$ and $\boldsymbol{C}(t)$.
Then, solve the following two subproblems in each time slot, namely
Joint Observation Scheduling and Adaptive Processing (JOSAP) subproblem
and Transmission Scheduling (TS) subproblem. A summary of DMRC algorithm
is given in Algorithm 1.
\begin{algorithm}
\caption{\label{alg:Dynamic-Multi-Resource}Dynamic Multi-Resource Cooperation
(DMRC) Algorithm}

\begin{algorithmic}[1]

\STATE At each time slot, observe $\boldsymbol{Q}(t)$, $\boldsymbol{P}(t)$,
$\boldsymbol{B}(t)$ and $\boldsymbol{C}(t)$.

\STATE Obtain the observation scheduling decision $\boldsymbol{X}(t)$
and compression ratio selection $\boldsymbol{\varrho}(t)$ by solving
the \textcolor{black}{joint observation scheduling and adaptive processing
subproblem (\ref{eq:subproblem_1}):}\vspace{-0.5em}
\begin{eqnarray}
 & \min_{\boldsymbol{X}(t),\boldsymbol{\varrho}(t)}\: & -V\sum_{i=1}^{I}U_{i}(A_{i}(t))+\sum_{i=1}^{I}\sum_{k=1}^{K}Q_{k}^{i}(t)A_{k}^{i}(t)-\sum_{i=1}^{I}P_{i}(t)A_{i}(t)\nonumber \\
 & s.t. & \textrm{C2},\textrm{C3},\textrm{C9},\textrm{C10}.\label{eq:subproblem_1}
\end{eqnarray}

\STATE Find transmission scheduling matrix $\boldsymbol{Y}(t)$ by
solving the transmission scheduling subproblem (\ref{eq:subproblem_2}):\vspace{-1.0em}
\begin{eqnarray}
 & \max_{\boldsymbol{Y}(t)}\: & \sum_{i=1}^{I}\sum_{k=1}^{K}Q_{k}^{i}(t)\mu_{k}^{i}(t)\nonumber \\
 & s.t. & \textrm{C4},\textrm{C5},\textrm{C6},\textrm{C8}.\label{eq:subproblem_2}
\end{eqnarray}

\STATE Update $\boldsymbol{Q}(t)$ and $\boldsymbol{P}(t)$ according
to (\ref{eq:queue_dynamic}) and (\ref{eq:virtual queue}), respectively.

\end{algorithmic}
\end{algorithm}

\subsubsection{\textcolor{black}{Joint observation scheduling and adaptive processing
subproblem}}

For the first subprobelm embedded in the proposed DMRC algorithm,
a lemma is introduced to transform the objective function in (\ref{eq:subproblem_1}).
\begin{lem}
\textup{Problem (\ref{eq:subproblem_1}) can be equivalently transformed
into the following problem, which is \vspace{-1.5em}}
\end{lem}
\begin{eqnarray}
 &  & \max_{\boldsymbol{X}(t),\boldsymbol{\varrho}(t)}\: V\sum_{i=1}^{I}\sum_{k=1}^{K}U_{i}(A_{k}^{i}(t))+\sum_{i=1}^{I}\sum_{k=1}^{K}Q_{k}^{i}(t)A_{k}^{i}(t)-\sum_{i=1}^{I}\sum_{k=1}^{K}P_{i}(t)A_{k}^{i}(t)\nonumber \\
 & s.t. & \textrm{C2},\textrm{C3},\textrm{C9},\textrm{C10}.\label{eq:sub1_trans}
\end{eqnarray}

\begin{IEEEproof}
See Appendix B.
\end{IEEEproof}
Observe that $A_{k}^{i}(t)=\varrho_{k}^{i}(t)x_{i,k}(t)B_{i,k}(t)$
is a nonlinear function in both $\varrho_{k}^{i}(t)$ and $x_{i,k}(t)$
in (\ref{eq:sub1_trans}). To tackle this difficulty, we can view
$A_{k}^{i}(t)$ as a separate decision variable for time slot $t$
and add corresponding constraints in the above optimization problem.
This yields \vspace{-1.0em}

\begin{eqnarray}
 &  & \max_{\boldsymbol{X}(t),\boldsymbol{\varrho}(t),\boldsymbol{A}(t)}\: V\sum_{i=1}^{I}\sum_{k=1}^{K}U_{i}(A_{k}^{i}(t))+\sum_{i=1}^{I}\sum_{k=1}^{K}Q_{k}^{i}(t)A_{k}^{i}(t)-\sum_{i=1}^{I}\sum_{k=1}^{K}P_{i}(t)A_{k}^{i}(t)\nonumber \\
 & s.t. & \textrm{C2},\textrm{C3},\textrm{C9},\textrm{C10}\nonumber \\
 &  & \textrm{C11}:\: A_{k}^{i}(t)\leq x_{i,k}(t)B_{i,k}(t)\nonumber \\
 &  & \textrm{C12}:\: A_{k}^{i}(t)\in\{0,\varrho_{1}B_{i,k}(t),...,\varrho_{J}B_{i,k}(t)\}.\label{eq:sub1_trans_2}
\end{eqnarray}
Note that $\varrho_{k}^{i}(t)$ is inherently captured in $A_{k}^{i}(t)$.
By relaxing discrete variables $x_{i,k}(t)$ and $A_{k}^{i}(t)$ into
continuous ones, modified problem (\ref{eq:sub1_trans_2}) turns into
a convex optimization problem with a concave objective function and
several linear inequalities constraints. The concavity of the objective
function can be verified via calculating its second derivative in
$A_{k}^{i}(t)$. To solve this convex problem, we first give the partial
Lagrangian of the primal problem (\ref{eq:sub1_trans_2}) as\vspace{-1.0em}

\begin{eqnarray}
 &  & L(\boldsymbol{X}(t),\boldsymbol{A}(t),\boldsymbol{\alpha}(t))=V\sum_{i=1}^{I}\sum_{k=1}^{K}U_{i}(A_{k}^{i}(t))+\sum_{i=1}^{I}\sum_{k=1}^{K}Q_{k}^{i}(t)A_{k}^{i}(t)\nonumber \\
 &  & -\sum_{i=1}^{I}\sum_{k=1}^{K}P_{i}(t)A_{k}^{i}(t)+\sum_{i=1}^{I}\sum_{k=1}^{K}\alpha_{i,k}(t)\left(x_{k}^{i}(t)B_{i,k}(t)-A_{k}^{i}(t)\right)\nonumber \\
 & s.t. & \textrm{C2},\textrm{C3},x_{i,k}(t)\in[0,1],A_{k}^{i}(t)\in[0,\varrho_{1}B_{i,k}(t)]\label{eq:Lagrange function}
\end{eqnarray}
where $\boldsymbol{\alpha}(t)=\{\alpha_{i,k}(t)\}$ are the set of
dual variables corresponding to C11. We then rearrange the terms of
(\ref{eq:Lagrange function}) as follows:\vspace{-1.0em}

\begin{eqnarray}
 &  & L(\boldsymbol{X}(t),\boldsymbol{A}(t),\boldsymbol{\alpha}(t))=L_{1}(\boldsymbol{A}(t),\boldsymbol{\alpha}(t))+L_{2}(\boldsymbol{X}(t),\boldsymbol{\alpha}(t))\nonumber \\
 &  & =\sum_{i=1}^{I}\sum_{k=1}^{K}\left\{ VU_{i}(A_{k}^{i}(t))+\left(Q_{k}^{i}(t)-P_{i}(t)-\alpha_{i,k}(t)\right)A_{k}^{i}(t)\right\} \label{eq:item1}\\
 &  & =+\alpha_{i,k}(t)\sum_{i=1}^{I}\sum_{k=1}^{K}\left(x_{k}^{i}(t)B_{i,k}(t)\right)\label{eq:item2}\\
 & s.t. & \textrm{C2},\textrm{C3},x_{i,k}(t)\in[0,1],A_{k}^{i}(t)\in[0,\varrho_{1}B_{i,k}(t)].\label{eq:Lagrange_Reformulation}
\end{eqnarray}

As such, if the optimal value of $\boldsymbol{\alpha}(t)$ is known,
the dual objective function $L(\boldsymbol{X}(t),\boldsymbol{A}(t),\boldsymbol{\alpha}(t))$
is decomposed into two independent parts $L_{1}(\boldsymbol{A}(t),\boldsymbol{\alpha}(t))$
and $L_{2}(\boldsymbol{X}(t),\boldsymbol{\alpha}(t))$, which correspond
to (\ref{eq:item1}) and (\ref{eq:item2}), respectively. On the basis,
problem (\ref{eq:Lagrange_Reformulation}) can be decomposed into
the following two subproblems:

\textit{1) Compression Ratio Selection}: The first one is the compression
ratio selection problem as follows\vspace{-1.0em}

\begin{eqnarray}
 &  & \max_{\boldsymbol{A}(t)}\: L_{1}(\boldsymbol{A}(t),\boldsymbol{\alpha}(t))\nonumber \\
 &  & s.t.\: A_{k}^{i}(t)\in[0,\varrho_{1}B_{i,k}(t)].
\end{eqnarray}

It can be solved using the standard optimization techniques and the
KKT conditions \cite{Book_ConvexOpt}. As a result, we derive the
optimal value $\left(A_{k}^{i}(t)\right)^{*}$ as \vspace{-1.0em}

\begin{equation}
\left(A_{k}^{i}(t)\right)^{*}=\begin{cases}
0, & \chi_{i,k}(t)\geq V\\
\varrho_{1}B_{i,k}(t), & \chi_{i,k}(t)\leq\frac{V}{1+\varrho_{1}B_{i,k}(t)}\\
\frac{V}{\chi_{i,k}(t)}-1, & \textrm{otherwise}
\end{cases}\label{eq:CRS_solution}
\end{equation}
where $\chi_{i,k}(t)=\alpha_{i,k}(t)+P_{i}(t)-Q_{k}^{i}(t)$. The
concrete solving procedure can be found in Appendix C. From (\ref{eq:CRS_solution}),
we can get the optimal compression ratio $\left(\varrho_{k}^{i}(t)\right)^{*}$
by projecting $\left(A_{k}^{i}(t)\right)^{*}/B_{i,k}(t)$ into the
set $\Lambda$.

\textit{2) Observation Scheduling}: \vspace{-1.0em}

\begin{eqnarray}
 &  & \max_{\boldsymbol{X}(t)}\: L_{2}(\boldsymbol{X}(t),\boldsymbol{\alpha}(t))\nonumber \\
 & s.t. & \textrm{C2},\textrm{C3},x_{i,k}(t)\in[0,1].\label{eq:observation scheduling}
\end{eqnarray}

In LP problem (\ref{eq:observation scheduling}), all constraints
are affine, and thus the feasible region is a polytope. According
to \cite{Book_ConvexOpt}, the optimal solution to an LP problem can
only be achieved at the extreme point. In what follows, we give a
lemma to show that each element within the optimal solution should
be binary.
\begin{lem}
\textup{\label{lem:binary}All elements $\{x_{i,k}^{*}(t)\}$ within
the optimal solution to problem (\ref{eq:observation scheduling})
are binary, i.e., $x_{i,k}^{*}(t)\in\{0,1\},\,\forall i,k$.}\end{lem}
\begin{IEEEproof}
We prove the lemma by contradiction. The detailed procedure is similar
to that in \cite{CL_SCMA}. Assume that there exists at least an element
$x_{\overline{i},\overline{k}}^{*}(t)$ within $\mathbf{x}=\{x_{i,k}^{*}(t)\}$
being non-binary, i.e., $0<x_{\overline{i},\overline{k}}^{*}(t)<1$.
On one hand, if constraints C2 and C3 are strict inequalities, we
can construct two points $\mathbf{x}_{1}=\{\cdots,x_{\overline{i},\overline{k}}^{*}(t)+\epsilon,\cdots\}$
and $\mathbf{x}_{2}=\{\cdots,x_{\overline{i},\overline{k}}^{*}(t)-\epsilon,\cdots\}$,
wherein $(\cdots)$ indicates that other elements are the same with
those in $\mathbf{x}$, and $\epsilon$ is an arbitrarily small positive
real number. As such, both $\mathbf{x}_{1}$ and $\mathbf{x}_{2}$
are within the polytope, and $0.5\mathbf{x}_{1}+0.5\mathbf{x}_{2}=\mathbf{x}$
holds. Hence, $\mathbf{x}$ is not an extreme point. On the other
hand, if at least one constraint in C2 and C3 is tight, i.e., the
equality holds, there must be another element $0<x_{\widehat{i,}\widehat{k}}^{*}(t)<1$
in $\mathbf{x}$ to guarantee the equality. Similarly, we can find
two points $\mathbf{x}_{1}=\{\cdots,x_{\overline{i},\overline{k}}^{*}(t)+\epsilon,x_{\widehat{i,}\widehat{k}}^{*}(t)-\epsilon,\cdots\}$
and $\mathbf{x}_{2}=\{\cdots,x_{\overline{i},\overline{k}}^{*}(t)-\epsilon,x_{\widehat{i,}\widehat{k}}^{*}(t)+\epsilon\cdots\}$
within the polytope, such that $0.5\mathbf{x}_{1}+0.5\mathbf{x}_{2}=\mathbf{x}$
holds as well. This verifies that $\mathbf{x}$ is not an extreme
point. To this end, Lemma \ref{lem:binary} is proved.
\end{IEEEproof}
From Lemma \ref{lem:binary}, we can equivalently transform problem
(\ref{eq:observation scheduling}) as \vspace{-1.0em}

\begin{eqnarray}
 &  & \max_{\boldsymbol{X}(t)}\: L_{2}(\boldsymbol{X}(t),\boldsymbol{\alpha}(t))\nonumber \\
 & s.t. & \textrm{C2},\textrm{C3},x_{i,k}(t)\in\{0,1\}.\label{eq:observation scheduling-1}
\end{eqnarray}

It can be seen that this problem is a weighted maximum matching problem
as discussed in the following, where the weight can be set as corresponding
observation link capacity. As a result, it can be solved optimally
using existing algorithms. We should emphasize that the observation
scheduling problem requires to be executed only once in each time
slot, regardless of $\boldsymbol{\alpha}(t)$ in each iteration for
solving the dual problem. This property greatly simplifies the solution
procedure.

After solving the two subproblems, we need to iteratively update the
Lagrange multipliers $\boldsymbol{\alpha}(t)$. For this aim, a subgradient
method can be used. More specifically, the $l$-th update can be performed
as\vspace{-1.0em}

\begin{equation}
\alpha_{i,k}^{l+1}(t)=\left[\alpha_{i,k}^{l}(t)-\lambda_{l}(x_{k}^{i}(t)B_{i,k}(t)-A_{k}^{i}(t))\right]^{+}
\end{equation}
where $\lambda_{l}$ is the sequence of scalar step sizes, and can
be set as any square summable but not absolute summable value.

\subsubsection{\textcolor{black}{Transmission scheduling subproblem}}

\begin{figure}
\centering\includegraphics[scale=0.5]{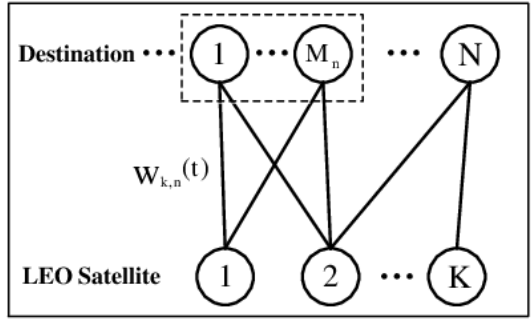}\vspace{-1.0em}

\caption{Bipartite graph representation for transmission scheduling problem.\label{fig:Bipartite-graph-representation.}}

\end{figure}

Subproblem (\ref{eq:subproblem_2}) can be reformulated into the following
problem:\vspace{-1.0em}

\begin{eqnarray}
 & \max_{\boldsymbol{Y}(t)}\: & \sum_{k=1}^{K}\sum_{d_{n}^{t}:(s_{k,\textrm{b}}^{t},d_{n}^{t})\in\mathcal{E}_{t}^{\textrm{tr}}}Q_{k}^{i^{*}}(t)y_{k,n}(t)C_{k,n}(t)\nonumber \\
 & s.t. & \textrm{C4},\textrm{C5},\textrm{C8}\label{eq:subproblem_2-reformulation}
\end{eqnarray}
where $i^{*}=\arg\max_{i}\, Q_{k}^{i},\,\forall k$. Note that $w(s_{k,\textrm{b}}^{t},d_{n}^{t},i)$
in (\ref{eq:subproblem_2-reformulation}) is replaced by $C_{k,n}(t)$
since it is optimal to serve the flow with maximum transmission link
capacity. To solve the reformulated problem, as shown in Fig. \ref{fig:Bipartite-graph-representation.},
we first utilize a bipartite graph to represent related components
in the SIN. The vertices therein can be divided into two disjoint
sets, i.e., the set of LEO satellites $\mathcal{K}$ and the set of
destinations $\mathcal{N}$. A destination $n$ is represented by
$M_{n}$ vertices. Accordingly, link weight $W_{k,n}(t)$ is selected
as the link capacity weighted queue backlog, $Q_{k}^{i^{*}}(t)C_{k,n}(t)$.
It can be observed that the formulated TS subproblem is identical
to the maximum weighted matching problem in the constructed bipartite
graph, and thereby efficient algorithms, e.g., Hungarian algorithm,
can be applied to obtain the optimal solution.

\subsection{Algorithm Analysis}

Herein, we first analyze the computational complexity of the proposed
DMRC algorithm. \textcolor{black}{The complexity of the DMRC algorithm
consists of two main parts: 2) JOSAP subproblem with the complexity
of $O(\frac{1}{\epsilon}+I^{3})$ --- It corresponds to complexity
of compression ratio selection problem $O(\frac{1}{\epsilon})$ \cite{TCOM06_DualMethods}
and complexity of observation scheduling problem $O(I^{3})$. For
the compression ratio selection problem, the subgradient method converges
to the desired state only after $O(\frac{1}{\epsilon})$ iterations,
where $\epsilon$ is the maximum tolerance deviation from the optimal
value. In each iteration, solving the problem only incurs $O(1)$
complexity, because a closed-form expression is available to the problem.
As for the observation scheduling problem, it takes complexity of
$O(I^{3})$ to solve the matching problem as described later. Thus,
the complexity for solving JOSAP subproblem is $O(\frac{1}{\epsilon})+O(I^{3})$;
2) Transmission scheduling subproblem with the complexity of $O(K^{3})$
--- This complexity comes from solving the maximum weighted matching
problem on the constructed bipartite graph. It takes $O\left(\left(\max\{K,\sum_{n}M_{n}\}\right)^{3}\right)\approx O\left(K^{3}\right)$
for }Hungarian algorithm\textcolor{black}{{} to obtain the optimal solution
\cite{Ref_HungarianMethod}. Note that we make the approximation since
the number of EOSs $K$ is generally larger than that of destinations
$\sum_{n}M_{n}$ (i.e., virtual vertices corresponding to all destinations);
Therefore, the overall computational complexity of DMRC algorithm
can be derived as $O(\frac{1}{\epsilon}+I^{3}+K^{3})$.}

In parallel, we present the performance results of the DMRC algorithm.
Specifically, the following theorem is given to characterize its asymptotic
queue backlog and network utility.
\begin{thm}
\textup{The proposed DMRC algorithm yields the following time average
queue backlog and network utility bounds:\vspace{-1.0em}}

\begin{equation}
\lim_{T\rightarrow\infty}\frac{1}{T}\sum_{t=1}^{T}\sum_{i=1}^{I}\sum_{k=1}^{K}\mathbb{E}\left\{ Q_{k}^{i}(t)\right\} \leq\frac{\Gamma+VU^{*}}{\delta_{1}}\vspace{-1.0em}
\end{equation}

\begin{equation}
\lim_{T\rightarrow\infty}\frac{1}{T}\sum_{t=1}^{T}\sum_{i=1}^{I}\mathbb{E}\left\{ U_{i}(A_{i}(t))\right\} \geq U^{*}-\frac{\Gamma}{V}.
\end{equation}
\end{thm}
\begin{IEEEproof}
See Appendix D.
\end{IEEEproof}
It can be noticed that Theorem 1 proves the utility optimality of
DMRC algorithm. Since Theorem 1 holds for all $V>0$, we can choose
a sufficiently large $V$ such that $\Gamma/V$ is arbitrarily small
and the achieved utility of DMRC algorithm is arbitrarily close to
optimal.

\section{Simulation Results\label{sec:Performance-Evaluation} }

In this section, extensive experiments are conducted using Matlab
simulator to evaluate the performance of our proposed DMRC algorithm.
We present the simulation results from two aspects. First, we evaluate
both the network utility and queue dynamics by changing the control
factor, $V$. Then, we investigate impacts of different network parameters
on the DMRC performance.

We have conducted a set of simulations of SINs operating at the time
horizon from 1 Jan. 2018 04:00:00 to 2 Jan. 2018 04:00:00. The time
horizon is \textcolor{black}{discretized} into 1440 equal time slots,
with the interval of each time slot being 60 seconds. Eight targets
are located on the Earth\textquoteright{}s surface, i.e., Mamiraua
($2^{\circ}$S, $66^{\circ}$W), Cape York ($11^{\circ}$S, $142.5^{\circ}$E),
Alaska Coast ($60^{\circ}$N, $148^{\circ}$W), Himalaya ($28^{\circ}$N,
$87^{\circ}$E), Sahara ($28^{\circ}$N, $11.5^{\circ}$E), Sumatra
($2^{\circ}$S, $103^{\circ}$E), Greenland ($69^{\circ}$N, $49^{\circ}$W)
and Bora ($16^{\circ}$S, $151^{\circ}$W). Meanwhile, a number of
LEO satellites are uniformly distributed over 2 sun-synchronous orbits
at a height of 619.6km with inclination $97.86^{\circ}$. We further
set two relay satellites locating at nominal longitudes of $176.76^{\circ}$E
and $16.65^{\circ}$E as the destinations. The set of available observation
links and transmission links are obtained using the Satellite Tool
Kit (STK) software. \textcolor{black}{The observation link capacity
is i.i.d. over time slots and randomly takes values from the set $\{600,800,1000\}$Mbps,
while the transmission link capacity is uniformly distributed in the
set $\{0,200,400\}$Mbps. The set of all compression ratio is assumed
to be $\{\frac{2}{3},\frac{1}{2},\frac{1}{3},\frac{1}{4}\}.$ }

\subsection{\textcolor{black}{Network Utility and Queue Dynamics}}

\begin{figure}
\centering\includegraphics[scale=0.6]{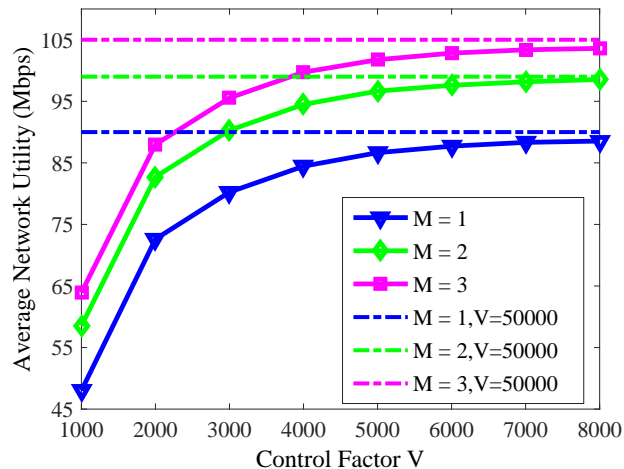}\vspace{-1.0em}

\caption{Average network utility versus control factor $V$.\label{fig:Average-network-utility-V}}
\end{figure}

In Fig. \ref{fig:Average-network-utility-V}, we evaluate the average
network utility performance versus the value of control factor $V$
ranging from 1000 to 8000. Clearly, the average network utility increases
with an increase of $V$. However, the rate of network utility increases
reduces with larger $V$. This is because the network utility is a
concave function of $V$. We take a large value of $V$ to illustrate
the optimal network utility ($V=50000$ in our setting). We compare
the network utility obtained by $V$ ranging from 1000 to 8000 to
the network utility obtained by $V=50000$. As shown in the figure,
the increase of network utility from $V=8000$ to $V=50000$ is quite
limited in comparison to the increasing from $V=1000$ to $V=8000$.
Therefore, the network utility achieved when $V=8000$ is close to
the value of optimal network utility.

\begin{figure}
\centering\includegraphics[scale=0.6]{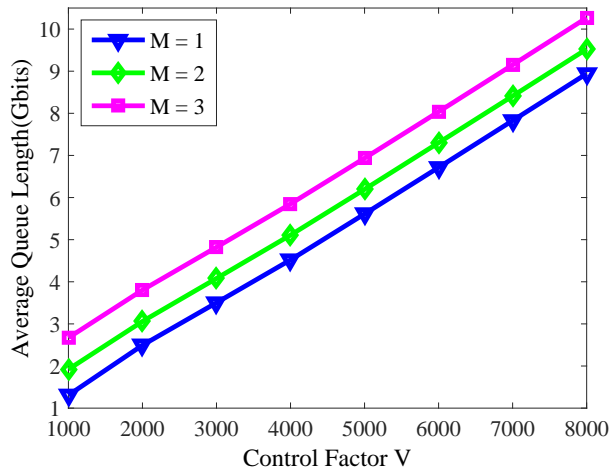}\vspace{-1.0em}

\caption{Average queue backlog versus control factor $V$.\label{fig:Average-queue-V}}
\end{figure}

Fig. \ref{fig:Average-queue-V} shows the data queue occupancy for
different values of $V$. The time-average lengths of data queues
increase with the value of $V$. It can be seen that, with the increase
of $V$, the network utility of DMRC algorithm asymptotically approaches
the optimality, which comes at a price of linearly growing average
queue length. The observation confirms the correctness of Theorem
1. On the other hand, we further observe that both the average network
throughput and queue backlog tend to increase with the growth in the
number of transceivers $M_{n}=M$, because more contacts can be established.

\subsection{\textcolor{black}{Performance Comparison}}

We validate the efficiency of proposed DMRC algorithm by comparing
it with two benchmark algorithms. The first one is random allocation
algorithm (denoted as Random), wherein observation resource, transmission
resource and compression ratio are randomly selected. The second one
is named as Fixed-CR algorithm, wherein both the observation and transmission
resources are allocated based on the maximum weighted matching method,
and the compression ratio is fixed at $\frac{1}{4}$. We set $V=8000$
and $M=2$ in the following simulations if not specified otherwise.

\subsubsection{The number of EOSs }

\begin{figure}
\centering\includegraphics[scale=0.6]{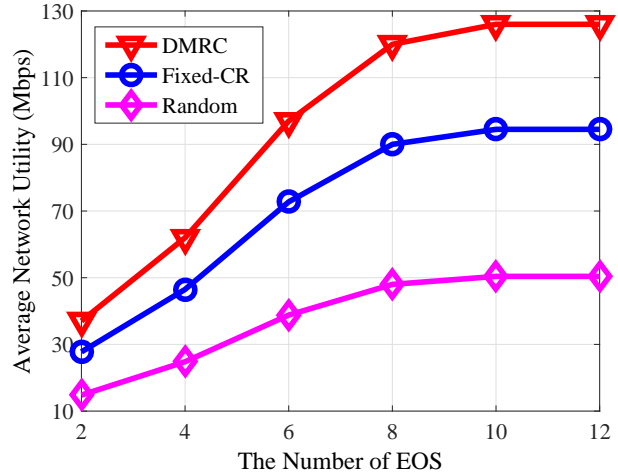}\vspace{-1.0em}

\caption{Average network utility versus the number of EOS.\label{fig:U_EOS_num}}
\end{figure}

In this experiment, we study the network performance of different
algorithms versus the number of EOSs. The minimum rate requirement
for an EOS is set to 0. Fig. \ref{fig:U_EOS_num} demonstrates that,
with the increase of the number of EOSs, the average network utility
improves for all the tested algorithms. As the number of EOSs grows,
there are more observation resources and transmission resources. Therefore,
each flow can get more chances to be served. It is observed that the
DMRC algorithm achieves higher network utility than that of Random
and Fixed-CR algorithms. The reason lies in that the DMRC algorithm
tends to build high quality observation and transmission links. Also,
the DMRC algorithm is able to adaptively select compression ratio
to process the acquired image data.

\begin{figure}
\centering\includegraphics[scale=0.6]{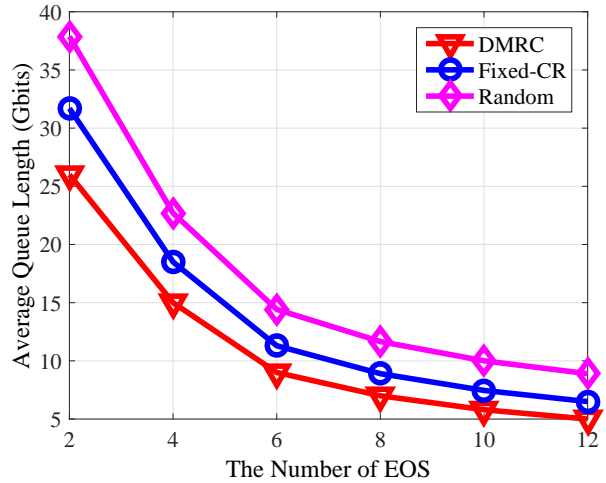}\vspace{-1.0em}

\caption{Average queue length versus the number of EOS.\label{fig:Q_EOS_num}}
\end{figure}

\begin{figure}
\centering\includegraphics[scale=0.6]{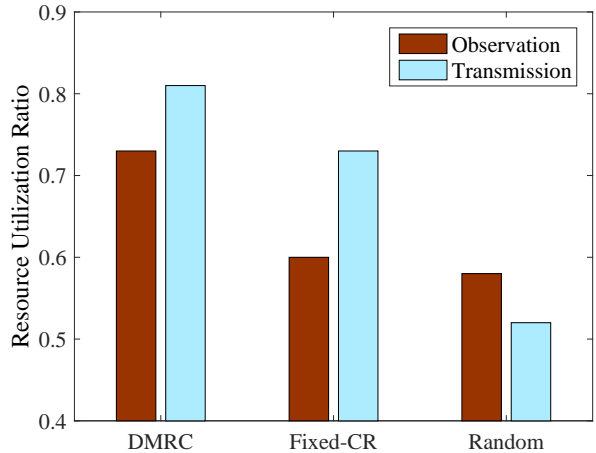}\vspace{-1.0em}

\caption{Resource utilization ratio for different algorithms.\label{fig:resource_utilization}}
\end{figure}

Fig. \ref{fig:Q_EOS_num} presents the average queue length performance
versus the number of EOSs. The average queue length decreases with
the increasing number of EOSs, as more resources can be utilized to
serve a flow. Besides, the DMRC algorithm yields the shortest queue
length. This can be explained for two aspects. On one hand, the DMRC
algorithm prefers to serve EOSs with longer queue length and adopts
higher compression ratio to reduce queue backlog. On the other hand,
through constructing high quality transmission links, the DMRC algorithm
can deliver observation data more quickly. To validate the performance
gain of the DMRC algorithm, we further plot the resource utilization
ratio for different algorithms. As depicted in Fig. \ref{fig:resource_utilization},
the proposed DMRC algorithm has the highest utilization ratio of both
the observation and transmission resources.

\subsubsection{Minimum data rate requirements}

\begin{figure}
\centering\includegraphics[scale=0.6]{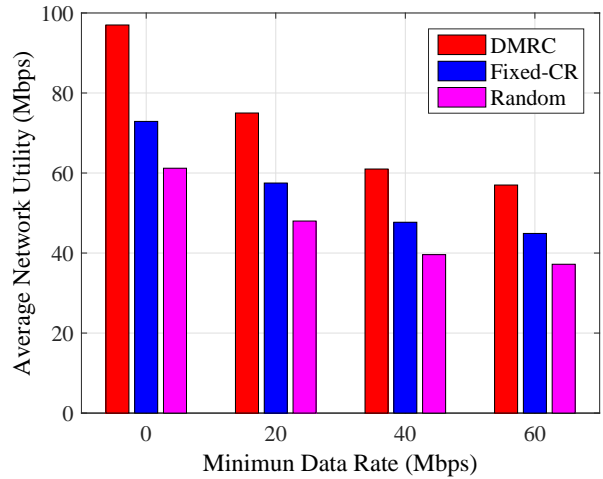}\vspace{-1.0em}

\caption{Average network utility versus the minimum data rate requirement.\label{fig:U_rate}}
\end{figure}

\begin{figure}
\centering\includegraphics[scale=0.6]{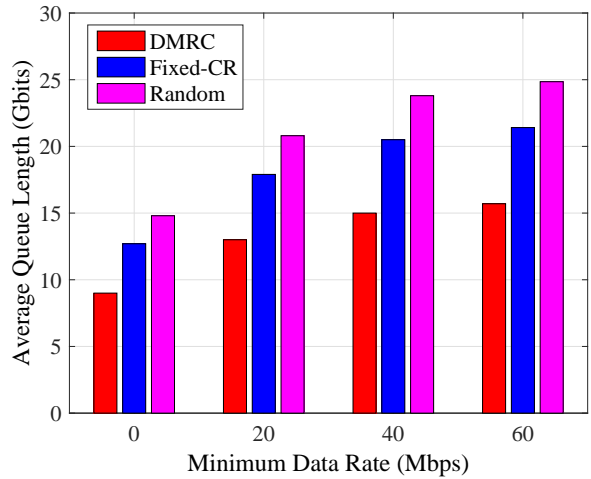}\vspace{-1.0em}

\caption{Average queue length versus the minimum data rate requirement.\label{fig:Q_rate}}
\end{figure}

We evaluate the effects of minimum required data rate requirements
on the network performance. In the following experiments, the number
of EOSs is equal to 6. As illustrated in Fig. \ref{fig:U_rate}, with
the minimum data rate varying from 0 to 60Mbps, the average network
utility for the three algorithms follows a declining trend. This can
be accounted by the fact that more resources are assigned to EOSs
with poor channel quality to ensure their minimum data rate requirements.
Hence, the network utility can be deteriorated to some extent. From
Fig. \ref{fig:Q_rate}, we observe that the average queue length becomes
larger with the rise of minimum data rate requirement. \textcolor{black}{The
trend is expected, because given a higher data rate, the bottleneck
EOSs need more resources to counteract undesired channel conditions.
This reduces the efficiency in resource allocation.} Another observation
is that the DMRC algorithm achieves superior performance with respect
to average network utility and queue backlog, in comparison with Random
and Fixed-CR algorithms, for the reason similar to that stated before.
\textcolor{black}{{} }

\subsubsection{The number of transceivers}

\begin{figure}
\centering\includegraphics[scale=0.6]{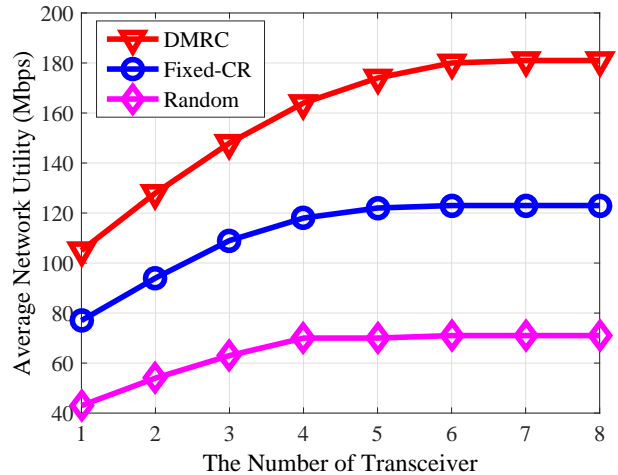}\vspace{-1.0em}

\caption{Average network utility versus the number of transceiver.\label{fig:U_trans}}
\end{figure}

\begin{figure}
\centering\includegraphics[scale=0.6]{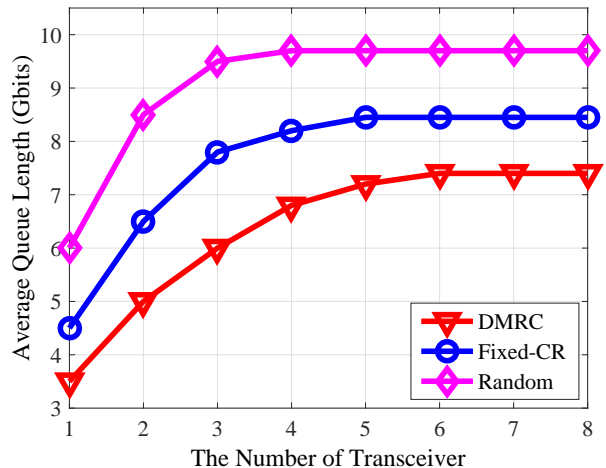}\vspace{-1.0em}

\caption{Average queue length versus the number of transceiver.\label{fig:Q_trans}}
\end{figure}

In the following simulations, we investigate the performance impacts
of the number of transceivers on destination relay satellites. The
number of EOSs is 12, and the minimum date rate requirement is zero.
Fig. \ref{fig:U_trans} and Fig. \ref{fig:Q_trans} depict the average
network utility and queue length performance for the three algorithms,
respectively. With an increasing number of transceivers, the average
network utility for DMRC, Random and Fixed-CR algorithms increases.
We attribute the results to more available transmission resources,
and thus more observation data can be transferred to destinations.
Moreover, a smaller compression ratio can be chosen to yield higher
image quality as well as network utility. However, as the number of
transceivers further increases, both the average network utility and
queue length remain steady. Under this circumstance, the observation
resource turns to become the bottleneck, and thus additional transmission
resource cannot be used for efficient data delivery.

\section{Conclusion\label{sec:Conclusion}}

\textcolor{black}{In this paper, we have investigated the problem
of multi-resource dynamic coordinate scheduling for SINs. Based on
the Lyapunov optimization theory, a low complexity algorithm, i.e.,
the DMRC algorithm, is proposed to properly balance among the observation,
computation and transmission resources. Extensive simulations have
been conducted to verify that the newly proposed DMRC algorithm performs
arbitrarily close to the optimal without requiring any prior knowledge.
}The impacts of various network parameters\textcolor{black}{{} are evaluated
on both the average network utility and queue backlog. There are a
few open issues to be addressed for future studies. First, we intend
to incorporate the reconfiguration delay into consideration when an
EOS switches between two observation targets. Second, we aim to exploit
the predictable network topology information to further reduce the
queue backlog of the proposed DMRC algorithm. Last but not the least,
how to effectively accommodate diverse QoS requirements, e.g., end-to-end
delay and observing resolution requirements, should be studied. }

\begin{appendices}  \section{Proof of Lemma 2}

According to the queue dynamics in (\ref{eq:queue_dynamic}), we have\vspace{-1.5em}

\begin{eqnarray}
 & Q_{k}^{i}(t+1)^{2}-Q_{k}^{i}(t)^{2}\nonumber \\
 &  & =\{[Q_{k}^{i}(t)-\mu_{k}^{i}(t)]^{+}+A_{k}^{i}(t)\}^{2}-Q_{k}^{i}(t)^{2}\nonumber \\
 &  & =\{[Q_{k}^{i}(t)-\mu_{k}^{i}(t)]^{+}\}^{2}+2A_{k}^{i}(t)[Q_{k}^{i}(t)-\mu_{k}^{i}(t)]^{+}+A_{k}^{i}(t)^{2}-Q_{k}^{i}(t)^{2}\nonumber \\
 &  & \leq\{Q_{k}^{i}(t)-\mu_{k}^{i}(t)\}^{2}+2A_{k}^{i}(t)Q_{k}^{i}(t)+A_{k}^{i}(t)^{2}-Q_{k}^{i}(t)^{2}\nonumber \\
 &  & =A_{k}^{i}(t)^{2}+\mu_{k}^{i}(t)^{2}+2Q_{k}^{i}(t)\{A_{k}^{i}(t)-\mu_{k}^{i}(t)\}.
\end{eqnarray}

Similarly, based on virtual queue formula (\ref{eq:virtual queue}),
we further obtain\vspace{-1.0em}

\begin{equation}
P_{i}(t+1)^{2}-P_{i}(t)^{2}\leq\{a_{i}-A_{i}(t)\}^{2}+2P_{i}(t)\{a_{i}-A_{i}(t)\}.
\end{equation}

Considering the boundness property of $A_{k}^{i}(t)$ and $\mu_{k}^{i}(t)$,
we can always find a constant, $\Gamma$, that satisfies \vspace{-1.0em}

\begin{equation}
\Gamma\geq\frac{1}{2}\sum_{i=1}^{I}\sum_{k=1}^{K}\{A_{k}^{i}(t)^{2}+\mu_{k}^{i}(t)^{2}\}+\frac{1}{2}\sum_{i=1}^{I}\{a_{i}-A_{i}(t)\}^{2}.
\end{equation}

Combining all the above three formulas, the following inequality holds\vspace{-1.0em}

\begin{equation}
L(\boldsymbol{\Theta}(t+1))-L(\boldsymbol{\Theta}(t))\leq\Gamma+\sum_{i=1}^{I}\sum_{k=1}^{K}\left\{ Q_{k}^{i}(t)(A_{k}^{i}(t)-\mu_{k}^{i}(t))\right\} +\sum_{i=1}^{I}\left\{ P_{i}(t)(a_{i}-A_{i}(t))\right\} .
\end{equation}

Adding $V\sum_{i=1}^{I}U_{i}(A_{i}(t))$ to both sides of the above
inequation and taking conditional expectations yield (\ref{eq:drift-plus-penalty}).
This completes the proof of Lemma 2.

\section{Proof of Lemma 3}

By substituting $A_{i}(t)=\sum_{k}A_{k}^{i}(t)$ into the objective
function, we have\vspace{-1.0em}

\begin{eqnarray}
 &  & \min\:-V\sum_{i=1}^{I}U_{i}(A_{i}(t))+\sum_{i=1}^{I}\sum_{k=1}^{K}Q_{k}^{i}(t)A_{k}^{i}(t)-\sum_{i=1}^{I}P_{i}(t)A_{i}(t)\nonumber \\
 &  & =\max\: V\sum_{i=1}^{I}U_{i}(\sum_{k=1}^{K}A_{k}^{i}(t))+\sum_{i=1}^{I}\sum_{k=1}^{K}Q_{k}^{i}(t)A_{k}^{i}(t)-\sum_{i=1}^{I}\sum_{k=1}^{K}P_{i}(t)A_{k}^{i}(t).\label{eq:sub1_obj}
\end{eqnarray}

Since $\sum_{k=1}^{K}x_{i,k}(t)\leq1$ and $x_{i,k}(t)\in\{0,1\}$
holds, there is at most one $k^{*}\in\mathcal{K}$ for flow $i$ such
that $x_{i,k^{*}}(t)=1$; Otherwise, $x_{i,k}(t)=0$. By extension,
there is at most one $k^{*}\in\mathcal{K}$ for flow $i$ such that
$A_{i,k^{*}}(t)>0$. To this end, we derive \vspace{-1.0em}

\begin{equation}
V\sum_{i=1}^{I}U_{i}(\sum_{k=1}^{K}A_{k}^{i}(t))=V\sum_{i=1}^{I}U_{i}(A_{k^{*}}^{i}(t))=V\sum_{i=1}^{I}\sum_{k=1}^{K}U_{i}(A_{k}^{i}(t)).\label{eq:obj_trans}
\end{equation}

By substituting (\ref{eq:obj_trans}) into (\ref{eq:sub1_obj}), we
obtain (\ref{eq:sub1_trans}). This completes the proof of Lemma 3.

\section{Calculation of Optimal Arrival Rate}

Recall that we set $U_{i}(A_{k}^{i}(t))=\log(1+A_{k}^{i}(t)),\:\forall k,i$
in this paper. Clearly, $U_{i}^{'}(A_{k}^{i}(t))=\frac{1}{1+A_{k}^{i}(t)}$.
Subsequently, we can get the first derivative of $L_{1}(\boldsymbol{A}(t),\boldsymbol{\alpha}(t))$
with respect to $A_{k}^{i}(t)$ as \vspace{-1.0em}

\begin{eqnarray}
 & L_{1}^{'}(\boldsymbol{A}(t),\boldsymbol{\alpha}(t)) & =\frac{V}{1+A_{k}^{i}(t)}-\left(\alpha_{i,k}(t)+P_{i}(t)-Q_{k}^{i}(t)\right)\nonumber \\
 &  & =\frac{V}{1+A_{k}^{i}(t)}-\chi_{i,k}(t).
\end{eqnarray}

\begin{enumerate}
\item If $L_{1}^{'}(\boldsymbol{A}(t),\boldsymbol{\alpha}(t))\leq0$ holds
for all $A_{k}^{i}(t)\in[0,\varrho_{1}B_{i,k}(t)]$, the maximum value
can be obtained at $A_{k}^{i}(t)=0$, because $L_{1}(\boldsymbol{A}(t),\boldsymbol{\alpha}(t))$
is a monotone decreasing function in the considered interval. In this
case, $\frac{V}{1+A_{k}^{i}(t)}-\chi_{i,k}(t)\leq0$ is always satisfied
within $[0,\varrho_{1}B_{i,k}(t)]$, which means $\chi_{i,k}(t)\geqslant V$.
\item If $L_{1}^{'}(\boldsymbol{A}(t),\boldsymbol{\alpha}(t))\geq0$ holds
for all $A_{k}^{i}(t)\in[0,\varrho_{1}B_{i,k}(t)]$, we can see that
$\chi_{i,k}(t)\leq\frac{V}{1+\varrho_{1}B_{i,k}(t)}$. Since $L_{1}(\boldsymbol{A}(t),\boldsymbol{\alpha}(t))$
is a monotone increasing function, the optimality is achieved at $A_{k}^{i}(t)=\varrho_{1}B_{i,k}(t)$.
\item Otherwise, the optimal value is gained by setting $L_{1}^{'}(\boldsymbol{A}(t),\boldsymbol{\alpha}(t))=0$.
This results in $A_{k}^{i}(t)=\frac{V}{\chi_{i,k}(t)}-1$.
\end{enumerate}
Summarizing the above cases, we finally attain the optimal solution
$\left(A_{k}^{i}(t)\right)^{*}$.

\section{Proof of Theorem 1}

We prove the theorem by comparing the Lyapunov drift with a stationary
and randomized algorithm denoted by $\Pi$. An algorithm $\Pi$ makes
a decision as a function of the observed state of the random events
in each time slot. It is independent of the queue backlog information.
According to the stochastic network optimization theory, given an
arbitrarily small $\epsilon>0$ and constant positive scalars $\delta_{1}$
and $\delta_{2}$, algorithm $\Pi$ can yield \vspace{-1.0em}

\begin{equation}
\mathbb{E}\left\{ U_{i}(\widetilde{A}_{i}(t))\right\} \leq U^{*}+\epsilon\vspace{-1.5em}\label{eq:e1}
\end{equation}

\begin{equation}
\mathbb{E}\left\{ \widetilde{A}_{k}^{i}(t)-\widetilde{\mu}_{k}^{i}(t)\right\} \leq\delta_{1}\vspace{-1.5em}\label{eq:e2}
\end{equation}

\begin{equation}
\mathbb{E}\left\{ a_{i}-\widetilde{A}_{i}(t)\right\} \leq\delta_{2}\label{eq:e3}
\end{equation}
where $\widetilde{A}_{k}^{i}(t)$ and $\widetilde{\mu}_{k}^{i}(t)$
are corresponding variables generated by algorithm $\Pi$. Since the
DMRC algorithm minimizes the right-hand-side of (\ref{eq:drift-plus-penalty}),
we have\vspace{-1.0em}

\[
\Delta(\boldsymbol{\Theta}(t))-V\sum_{i}\mathbb{E}\left\{ U_{i}(A_{i}(t))|\boldsymbol{\Theta}(t)\right\} \leq\Gamma-V\sum_{i=1}^{I}\mathbb{E}\left\{ U_{i}(\widetilde{A}_{i}(t))|\boldsymbol{\Theta}(t)\right\} \vspace{-1.5em}
\]

\begin{equation}
+\sum_{i=1}^{I}\sum_{k=1}^{K}\mathbb{E}\left\{ Q_{k}^{i}(t)(\widetilde{A}_{k}^{i}(t)-\widetilde{\mu}_{k}^{i}(t))|\boldsymbol{\Theta}(t)\right\} +\sum_{i=1}^{I}\mathbb{E}\left\{ P_{i}(t)(a_{i}-\widetilde{A}_{i}(t))|\boldsymbol{\Theta}(t)\right\} .\label{eq:e4}
\end{equation}

Substituting (\ref{eq:e1})\textendash{}(\ref{eq:e3}) into (\ref{eq:e4}),
we get the following inequation as $\epsilon\rightarrow0$ \vspace{-1.0em}

\begin{equation}
\Delta(\boldsymbol{\Theta}(t))-V\sum_{i}\mathbb{E}\left\{ U_{i}(A_{i}(t))|\boldsymbol{\Theta}(t)\right\} \leq\Gamma-VU^{*}+\delta_{1}\sum_{i=1}^{I}\sum_{k=1}^{K}\mathbb{E}\left\{ Q_{k}^{i}(t)\right\} .\label{eq:e5}
\end{equation}

Taking iterated expectation of (\ref{eq:e5}) over $t\in\{1,...,T-1,T\}$
and using the law of telescoping sums result in \vspace{-1.0em}

\[
\mathbb{E}\left\{ L(\boldsymbol{\Theta}(T))\right\} -\mathbb{E}\left\{ L(\boldsymbol{\Theta}(0))\right\} -V\sum_{t=1}^{T}\sum_{i=1}^{I}\mathbb{E}\left\{ U_{i}(A_{i}(t))\right\} \vspace{-1.5em}
\]

\[
\leq T\left(\Gamma-VU^{*}\right)+\delta_{1}\sum_{t=1}^{T}\sum_{i=1}^{I}\sum_{k=1}^{K}\mathbb{E}\left\{ Q_{k}^{i}(t)\right\} .
\]

Reducing the left-hand-side, increasing its right-hand-side by eliminating
all negative terms, and then rearranging it, we get\vspace{-1.0em}

\[
\lim_{T\rightarrow\infty}\frac{1}{T}\sum_{t=1}^{T}\sum_{i=1}^{I}\sum_{k=1}^{K}\mathbb{E}\left\{ Q_{k}^{i}(t)\right\} \leq\frac{\Gamma+VU^{*}}{\delta_{1}}.
\]

Similarly, we can obtain \vspace{-1.0em}

\[
\lim_{T\rightarrow\infty}\frac{1}{T}\sum_{t=1}^{T}\sum_{i=1}^{I}\mathbb{E}\left\{ U_{i}(A_{i}(t))\right\} \geq U^{*}-\frac{\Gamma}{V}.
\]

\end{appendices}

\bibliographystyle{IEEEtran}
\bibliography{ref_satellite_1,ref_satellite,tex}

\end{document}